\documentclass[11pt]{article}
\usepackage{mathtools}
\usepackage{upgreek} 
\usepackage[utf8]{inputenc}
\usepackage[T2A]{fontenc}
\usepackage[ukrainian,english]{babel}
\title{\textbf{ATMOSPHERIC AEROSOL LIGHT SCATTERING SPECTRAL AND POLARIZATION PECULIARITIES}}
\author{Zh.I.Patlashenko \\ \\ State Ecological Academy \\ of Post-graduate Education and Management,\\ Kyiv,
Ukraine, deazh@ukr.net}
\date{}
\begin{document}

\maketitle

\begin{abstract}
This paper considers environmental problems of natural and anthropogenic atmospheric aerosol pollution and its global and regional monitoring. Efficient aerosol investigations may be achieved by spectropolarimetric measurements. Specifically second and fourth Stokes parameters spectral dependencies carry information on averaged refraction and absorption indexes and on particles size distribution functions characteristics.

\end{abstract}

\section{INTRODUCTION}

Light scattering in the Earth atmosphere is a complex process that depends on different atmospheric components concentration and state and also on averaged parameters, first of all on temperature and pressure of the atmosphere specific layer.

Atmospheric aerosol is an inseparable atmospheric compound that influences its ecological and climate state. Small aerosol particles sized from a few nanometers to a few micrometers have diverse compound, structure and physical-and-chemical properties \cite{1,2,3,4}. Despite the multiplicity of aerosol interaction with light quants it is quiet amenable to regional and seasonal modeling under normal conditions. Such models are used as a basis for remote optical investigation methods.

However, in case of non-regular natural or technogenic aerosol emissions, such as volcano eruptions, hurricanes, sand storms, anthropogenic point and surface sources, etc, most of contemporary models do not provide for sufficient quality of data interpretation, often falling back to some non-informative values like aerosol optical absorbency. This problem is exceptionally important in case of global monitoring, when monitoring data is interpreted based on high-scale models that do not consider local peculiarities.

\section{Aerosol environmental impact}

Atmospheric aerosol subcomponents can be natural and anthropogenic. Naturally aerosols form due to diverse natural physical and chemical phenomenon, such as evaporation and condensation, turbulent gas and hydrodynamic processes, photochemical and chemical reactions, etc. \cite{5}. Aerosol is also a result of anthropogenic activities like production, chemical, mining, building industries, burning of oil products, coal, gas, etc. Another important aerosol formation factor is aerosol lifting by wind due to soil erosion, which contains natural and/or anthropogenic particles. The largest aerosol quantity is sea aerosol, and anthropogenic aerosol constitutes around 10\% of total aerosol mass \cite{6}.

Atmospheric aerosol influences global climate, changing the amount of solar radiation that reaches the Earth surface, and influencing atmospheric transparency in Earth infrared emission band. These phenomenon are stipulated by a set of direct, indirect and semi-indirect aerosol effects \cite{7,8,9,10}.

Direct aerosol effect is caused by absorption and scattering of light by atmospheric aerosol. Indirect aerosol effect is caused by Earth albedo change due to altering optical and physical clouds characteristics. aerosol as condensation nuclei increases the number of raindrops (first indirect effect or Twomey effect \cite{8,11,12,13}) and increases their volume, delaying rainfall and prolonging the cloud lifetime (second indirect effect or Albrecht effect \cite{9,12,13,14}).

Semi-direct aerosol effect includes all effects that do not fit into direct and indirect aerosol effects definition, such as moisture height redistribution \cite{15}, vertical moisture fluxes stabilization \cite{16}, vertical temperature profile change \cite{9}, etc.

An important ecological phenomenon is atmospheric aerosol deposition that changes Earth surface albedo. An example of this effect is anthropogenic aerosol deposition on Arctic ice that causes its melting due to albedo decrease.

Besides changing climate aerosol also directly influences biosphere objects and eventually directly or indirectly affects humans. Since 1970s many mortalities are linked to anthropogenic atmosphere aerosol pollution \cite{17,18,19}. Yearly death rate in USA linked to atmospheric aerosol is estimated to be 22 to 52 thousands \cite{20}. And total quantity of respiratory diseases due to aerosol pollution increases by 50\% every 5 years \cite{21}.

It is well known that inhalation of small aerosol particles may cause asthma, lungs cancer, cardiovascular diseases, respiratory diseases, birth defects and premature death.

Large aerosol particles are mostly filtered by epipharynx. However, particles with size under 10 micrometers may be transported deeply into lungs - to bronchi and alveoli. These particles are regulated by PM10 standard in EU \cite{22}

Particles under 2.5 micrometers are regulated by PM2.5 ecological standard and may be transported to lungs gas exchange areas causing plaques in arteries, inflammation of blood vessels, arteriosclerosis, and other cardiovascular diseases \cite{23}. According to World Health Organization estimates PM2.5 particles cause around 3\% mortality due to cardiopulmonary diseases, 5\% mortality due to trachea, bronchi and lungs cancer and 1\% mortality due to acute respiratory infection in children under 5 years old \cite{17}. Moreover, such effects may appear even due to short-term inhalation of corresponding aerosol particles \cite{25}.

Aerosol particles under 100 nanometers, like exhausts of contemporary “clean” diesel engines, freely pass through lungs to blood and directly disorder internal organs including brain. This aerosol can carry carcinogenic compounds such as benzopyrenes.

Aerosol particles permeability to human organs is determined not only by their size, but also by their shape and physical-and-chemical compound \cite{26}. The problem of atmospheric aerosol ecological hazard dependency on aerosol shape is underexplored. Only general considerations that “sharp edged” aerosol particles (like asbestos) are more dangerous than “smooth edged” particles are formulated. Nanoscale aerosol particles that have increased surface area in comparison to spherical particles have higher chance of accumulating different hazardous substances at their surface.

Let's stress, that it is incorrect to estimate total mass of atmospheric aerosol contamination as an ecological parameter or standard, because a single 10 micron particle is much more safer than a 1000 of 100 nm particles that have 100 times less total mass. In some countries a total aerosol surface area regulation is proposed.

Another source of environmental hazard of natural and anthropogenic aerosol pollution is indirect impact at human health through food products of animal and vegetation origin, and through damage of natural ecosystems. E.g. increased atmospheric aerosol concentration may cause death of some plants \cite{27}.

Moreover, aerosol may pose direct hazard to health of living organisms, e.g. highly toxic aerosol or bacteria and viruses. Particularly there is a progressive problem of highly hazardous radioactive atmospheric aerosol represented by 0.02 to 1 micron particles, e.g. due to nuclear accidents similar to Chernobyl and Fukushima-Daiichi. Radioactive aerosols are referenced to as “low-activity” (radioactivity below $10^{-13}$ Ci), “semi-hot” (radioactivity from $10^{-13}$ to $10^{-10}$ Ci) and “hot” (radioactivity over $10^{-10}$ Ci). According to the source they are separated into natural, explosive (formed by nuclear device detonation) tests and industrial formed during nuclear substances management.

Radioactive aerosol inhalation is much more hazardous for living organisms than equivalent external irradiation as they can penetrate into the body and create internal irradiation which directly influences internal organs by focal necrosis. Only 10 to 50\% of radioactive aerosols can be efficiently removed from the body. Average radioactive aerosol troposphere suspension time varies from 2 to 30 days depending on weather and local atmospheric peculiarities and may be quickly transported around the globe.

\section{Contemporary state of atmospheric aerosol monitoring}

Being aware of natural and anthropogenic aerosol hazard the world community and individual countries pay a lot of attention to hold constant monitoring of atmosphere aerosol contamination, especially during extreme events of technological or natural origin. All atmospheric aerosol monitoring methods are strictly separated in two classes: contact probing and remote sounding.

Local survey is made near the point pollutant objects \cite{28}, around perimeter of area pollution sources \cite{29} and near population centers or special ecological control areas.

Contact probing is made by special hydrometeorological laboratories by analyzing chemical compound of atmospheric precipitation or atmospheric air probes and include aerosol mass spectrometry, differential mobility analysis, aerodynamic particle sizing, wide range particle spectrometry, micro-orifice uniform deposit impactors, condensation particle counters, epiphaniometry, electronic microscopy, instrumental neutron-activation analysis, etc.

On one hand contact probing is the most accurate and reliable atmospheric aerosol ecological state investigation method, that provides for their implementation in international and national law. On the other hand they are extremely confined spatially.

Non-contact remote sounding methods provide for lower accuracy and reliability in comparison to contact probing laboratory analysis. However they enable implementation of cost-efficient atmospheric aerosol parameters estimate. Classical remote sounding methods include solar photometers, polarimeters and lidars that provide for aerosol optical density determination. In some cases aerosol particles size distribution function maximum can also be determined.

Remote sounding methods are separated into active and passive. Active methods include lidars investigating laser emission scattering by atmosphere. It provides for aerosol vertical distribution estimation and does not depend on phase angle. Obviously such method requires rather complex infrastructure.

There is a separate class of indirect sounding methods that monitor specific aerosol type influence at different atmospheric components or underlying surface. A good example is air ionization by radioactive aerosols detection by standard military or civil radars providing for quantitative estimate of its emission and transfer.

Passive methods investigate solar light scattered by Earth atmosphere. Therefore such methods are influenced by additional uncertainty linked to solar radiation parameters variation. Phase angle temporal variation and different atmospheric components spectra overlap due to rescattering, luminescence, reabsorption of solar light, etc introduce additional errors.

Partially those problems can be solved by direct solar calibration \cite{30}, most efficient in space conditions, and also by multifrequency spectrometry or spectrophotometry also providing for more extensive atmosphere investigation.

Remote sounding methods may be ground or airplane/satellite based. Due to increased complexity and ambiguousness of satellite data interpretation the ground stations data that covers only relatively small survey area is used as validation for global satellite monitoring.

Therefore current state of atmospheric aerosol monitoring does not solve the problem of high-quality global monitoring. Ground-based contact and remote investigation stations despite their accuracy are limited locally and satellite methods lack accuracy and reliability and are limited by measured parameters quantity.

A quality step in optical remote sounding methods is detection of spectral and polarization peculiarities of the signal that would increase the amount of the experimental data obtained 4 times with minimal information efficiency loss.

\section{Problems of atmospheric aerosol scattering matrix remote monitoring}

Aerosol particles interaction with light quants is significantly different from gas atoms and molecules which differs first of all for polarization properties of the scattered light.

The scattering matrix is a quasi-symmetric Muller matrix constituting 16 components and describe transformation of incident light Stokes vector to scattered light Stokes vector \cite{31}. Specific values of scattering matrix components depend on averaged atmospheric aerosol characteristics and their dispersion and also on orientation of anisotropic particles relative to incident light. In some cases the matrix form is simplified by symmetry.

Determination of all independent scattering matrix components is possible only in case there is a way to control the incident light polarization. And determination of specific aerosol properties inverse problem may be solved only in case scattering matrix components are known for a wide set of phase angles.

Active Stokes-polarimetry may be implemented at autonomous probes in the Earth atmosphere \cite{32}, however, limitation of such methods are the same as for contact probing. In remote sounding lidars provide for both incident light polarization control and vertical distribution determination.

On the other hand, lidars have their engineer and technical limitations both from size and mass and from power consumption perspective. Moreover, reverse scattered by most atmospheric aerosol light polarization is nearly identical to that of incident light, therefore a synchronous observation system must be established of spatially separated lasers and detectors, limiting system mobility and efficiency (synchronous ground-and-space systems are also possible e.g. to validate satellite monitoring data). Further, even in case of a limited amount of spatially-separated detectors is present, the polarization is still determined at a discrete set of phase angles that reduces efficiency of aerosol parameters determination by their comparison with the model. And finally, there are ecological problems in operating wavelength-tunable lasers that contain environmentally hazardous organic dyes and acids.

On the other hand passive devices can be relatively compact to be used in a space vehicle. At present only spectrometers (measuring only 1 Stokes parameters) and photopolarimeters (measuring 2 and more Stokes parameters in a wide spectral bandwidth) are used failing to separate gas and aerosol phenomena in the atmosphere and to cancel out underlying Earth surface influence. Therefore there is no contemporary efficient method of atmospheric aerosol scattering matrix monitoring.

\section{Aerosol light scattering peculiarities}

Aerosol light scattering is usually separated in two different cases: small particles (particle size smaller than the wavelength) and large particles, which are usually considered spherical. However, spherical particles are not observed in nature and even a raindrop has aspherical shape due to gravitation forcing and gas-dynamic processes. Crystals, snowflakes, smoke particles and dust formed by minerals fragmentation surface shape is very complex and scattering matrix model determination is a complex theoretical task \cite{34}. Rough particles are a separate class of aerosols, which have surface inhomogeneities size comparative to the wavelength \cite{35}.

It was experimentally determined that spherical atmospheric aerosol model (Mie theory) is not correct for chaotic oriented fractured glass particles \cite{36} and false phase dependencies of matrix components are obtained. Significant discrepancies in linear polarization degree of crystal ammonia formed at temperatures from 130 to 180 $ ^{o}$K at wavelengths 470, 652 and 937 nm were also found experimentally \cite{37}.

Main discrepancies between spherical and chaotic oriented aspherical particles are following: spherical particles have more intense interference structures in second and third Stokes parameters and scattering indicatrix phase dependencies and higher reverse scattering intensity. However, total single-scattering albedo does not significantly depend on particles shape peculiarities \cite{33}.

Second and third Stokes parameters value and indicatrix for large particles strongly depends on refraction index and size distribution function parameters. Frequency and amplitude of the polarization degree oscillations relative to phase angle, particle size and refraction index are larger for monodisperse aerosol and smaller for polydisperse aerosol. Moreover, direct and reverse scattering intensity increased with increase of size dispersion, while reverse scattered light intensity is proportional to refraction index.

We should also note that the aerosol size is straightforward only for spherical particles. In case of aspherical particles “equivalent volume” (identical volume sphere) and “spherical shape factor” (sphere to particle surface area ratio) are introduced. Sometimes “sedimentary” or “Stokes” radius is also determined being the sphere radius with equal sedimentation speed.

\section{Off-atmospheric aerosol remote spectropolarimetric monitoring}

\begin{figure}
\begin{center}
\includegraphics[width=0.6\textwidth]{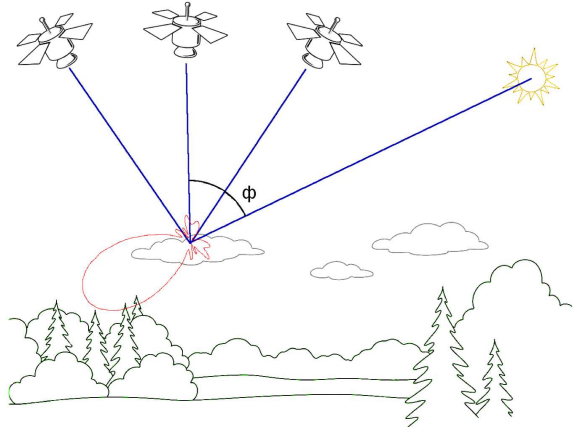}
\vskip-3mm\caption{Atmospheric aerosol monitoring geometry, $\upphi$ - observation phase angle, i.e. the angle of the light ray from the source to the observed volume and the light ray scattered in the observer direction.}
\label{fig:fig1}
\end{center}
\end{figure}

One of the most promising methods for refraction index determination is phase dependency of second and third Stokes parameters determination (fig.~\ref{fig:fig1}). Its spectral dispersion may also be used in case phase angle spread is limited.

Given that polarization degree is formed in upper atmosphere layers \cite{33}, its vertical inhomogeneity and particles vertical stratification may be neglected. In case absorption index is less than $10^{-3}$, the second and third Stokes parameters are determined mainly by refraction index, which is true for most natural atmospheric aerosols.

However, diversity of forms and structure of polydisperse aerosol ensemble is hard-to-model and may significantly influence monitoring data interpretation.

Polarization degree phase and spectral dependencies analysis provides for simultaneous determination of refraction index, maximum and dispersion of size distribution function, and optical absorbency by comparison with the model. First of all this is true for wavelengths over 500nm where atmospheric gas component influence may be neglected.

Second and third Stokes parameters analysis in a wide set of phase angles provides for the most accurate monitoring data. However, practically their spectral dependency may be measured instead under assumption that refraction index dispersion is negligibly small or known, which is approximately true for many types of aerosols of different origin in visible light.

There is also a method for atmospheric aerosol parameters determination by measuring fourth Stokes parameter \cite{38}, which is determined only by multiple scattering. The obtained aerosol parameters are averaged by a larger atmosphere depth and can be different than that obtained by second Stokes parameter measurement. Therefore the difference carries information on aerosol vertical distribution.

Absorption index can be estimated by aerosol single-scattering albedo spectral dependency \cite{39}. However, such method has a high error due to neglecting aerosol vertical distribution and strong dependency on aerosol particles shape model. E.g. in case of sub-micron particles the discrepancy was estimated to be up to 100\% \cite{40}.

\section{CONCLUSIONS}

1.	Investigations of scattering matrix and atmospheric aerosol parameters by spectropolarimetry may be divided into two types: one-channel observations at different phase angles and multichannel observations at a fixed phase angle.

2.	While Stokes parameter determined at different phase angles provides for higher estimate accuracy, this method does not provide for highly dynamic processes investigation and hinders global monitoring implementation.

3.	Refraction and absorption indexes, main parameters of aerosol size distribution functions and aerosol optical absorbency can be estimated based on backscattered light second and third Stokes parameters spectral and phase dependencies.

4.	This data may be validated and adjusted based on 4th Stokes parameter monitoring. Moreover, some conclusions on its vertical distribution can also be made.

\end{document}